\documentclass[journal]{IEEEtran}

\usepackage{amsmath,amssymb,amsfonts,amsthm,epsfig,euscript,graphicx}

\newcommand{\pr}{{\mathbb{P}}}
\newcommand{\ex}{{\mathbb{E}}}
\newcommand{\E}{{\EuScript{E}}}
\newcommand{\openone}{\leavevmode\hbox{\small1\normalsize\kern-.33em1}}

\newcommand{\etas}{{\eta^\circledast_{\ell}}}
\newcommand{\etans}{{\eta^\star_{\ell}}}
\newcommand{\e}{{\mathtt{e}}}

\newcommand{\tl}{{\tau_\ell}}
\newcommand{\eps}{{\varepsilon}}

\newcommand{\defeq}{\overset{\text{def}}{=}}

\newtheorem{lem}{Lemma}

\newtheorem{thm}{Theorem}
\newtheorem{fact}{Fact}

\begin{document}

\title{Estimating a Random Walk
First-Passage Time  from Noisy or Delayed
Observations}
\author{Marat V. Burnashev and Aslan
Tchamkerten,~\IEEEmembership{Member,~IEEE}\thanks{This work was supported in
part by the Russian Fund for Fundamental
Research (project number 09-01-00536) and an Excellence Chair Grant from the French National Research Agency (ACE
project).

M. V. Burnashev is with the
Institute for Information Transmission Problems, Russian Academy of Sciences
Moscow, Russia. A. Tchamkerten is with Telecom ParisTech,
Paris, France. Email: burn@iitp.ru, aslan.tchamkerten@telecom-paristech.fr }}

\maketitle \begin{abstract} A random walk (or a Wiener process), possibly with
drift, is observed in a noisy or delayed fashion. The problem considered in
this paper is to estimate the  first time $\tau$ the random walk reaches a
given level. Specifically, the $p$-moment ($p\geq 1$) optimization problem
$\inf_\eta \ex|\eta-\tau|^p$ is investigated where the infimum is taken over
the set of stopping times that are defined on the observation process.

When there is no drift, optimal stopping rules are characterized
for both types of observations. When there is a drift, upper and
lower bounds on $\inf_\eta \ex|\eta-\tau|^p$ are established for both types of
observations. The bounds are tight in the large-level regime for noisy
observations and in the large-level-large-delay regime for delayed
observations. Noteworthy, for noisy observations there exists an
asymptotically optimal stopping rule that is a function of a single
observation. 

Simulation results are provided that corroborate the validity of the results
for non-asymptotic settings. \end{abstract}

\section{Introduction}Suppose $X=\{X_t\}_{t\geq 0}$ is a stochastic process and
$\tau$ a stopping time defined over $X$.\footnote{Recall that a stopping time
with respect to a stochastic process $\{X_t\}_{t\geq 0}$ is a random variable
$\tau$ taking on values in the positive integers such that $\{\tau=t\}\in
{\cal{F}}_t$, for all $t \geq 0$, where ${\cal{F}}_t$ denotes the
$\sigma$-algebra generated by $X_{0},X_1,\ldots,X_{t}$.} Statistician has
access to $X$ only through correlated observations $Y=\{Y_t\}_{t\geq 0}$ and 
wishes to find a stopping $\eta$ defined over $Y$ that gets as close as
possible to $\tau$, for instance, so as to minimize some average absolute
moment $\ex|\eta-\tau|^p$. This general formulation was introduced in \cite{NT} as the
Tracking Stopping Time (TST) problem, and an early instance of it
where $Y=X$ and where $\tau$ is a randomized stopping time was investigated in
\cite{Mo2}.

The TST
problem generalizes the long studied Bayesian change-point detection
problem (see, e.g., \cite{TM} and the books \cite{PH} and \cite{BN2} for surveys on theory
and applications of the change-point problem). 

In the Bayesian change-point problem, there is a random variable
$\theta$, taking on values in the positive integers, and two
probability distributions $P_0$, the ``nominal'' distributions, and $P_1$, the
``alternative'' distribution. Under $P_0$, the
conditional density function of $Y_t$ given
$Y_0,Y_1,\ldots,Y_{t-1}$ is $f_0(Y_t|Y_0,Y_2,\ldots,Y_{t-1})$, for
every $t\geq 0$. Under $P_1$, the conditional density function of
$Y_t$ given $Y_0,Y_1,\ldots,Y_{t-1}$ is
$f_1(Y_t|Y_0,Y_1,\ldots,Y_{t-1})$, for every $t\geq 0$. The
observed process is distributed according $P_\theta$, which assigns
the conditional density functions of $P_0$ for all
$t<\theta$, and the conditional density functions of $P_1$
for all $t\geq \theta$. The Bayesian change-point problem typically consists in finding a
stopping time $\eta$, with respect to $\{Y_t\}$, that minimizes
some (loss) function of the delay $\eta -\theta$. 

To see that the Bayesian change-point problem can always be formulated as a TST
problem, it suffices to define the process
$X=\{X_t\}_{t\geq 0}$ as $X_t=0$ for $t<\theta$ and  $X_t=1$ for $t\geq \theta$. The Bayesian
change-point problem becomes the TST problem which consists in tracking $\theta$
(now defined as a stopping time with respect to $X$) through $Y$. 

The difference between the Bayesian change-point problem and the TST problem
lies in the equality
$$\pr(\theta=k|\tau>n, y^n)=\pr(\theta=k|\tau>n)\qquad k>n$$ which always holds
for the former but need not hold for the latter~\cite{NT}. In other
words,  for TST
problems past observations are in general useful for
estimating the future value of $\tau$, by contrast with Bayesian
change-point problems. For specific applications of the TST
problem formulation related to monitoring, communication, and forecasting
we refer to \cite[Section I]{NT}.

In \cite{NT}, through a computer science approach, a general
algorithmic solution is proposed for constructing optimal
``trackers'' for the cases where $X$ and $Y$ are processes defined
over finite alphabets and $\tau$ is bounded.  What motivated an
algorithmic approach is that the TST problem generalizes the
Bayesian change-point  problem for which  general closed-form
analytical solutions have been reported only for specific asymptotic
regimes, typically the vanishing false-alarm regime (see, e.g.,
\cite{Lai1}). Non-asymptotic closed-form solutions have been obtained essentially
for i.i.d. cases where, conditioned on the change-point value,
observations are independent with common distribution $P_0$ and $P_1$
before and after the change, respectively (see, e.g.,
\cite{Shi3,Shi}).\footnote{An exception is \cite{Y} which considers
Markov chain distributions, but of finite state.} 

Two natural TST settings include the ones where the observation process $Y$ is
a noisy or delayed version of $X$. In this paper we investigate both
situations when $X$ is a Gaussian random walk (or a Wiener process) possibly
with drift, and $\tau$ is the first time when $X$ reaches some given level
$\ell$. For noisy and delayed observations, we establish lower bounds on
$$\inf_\eta\ex|\eta-\tau|^p \qquad p\geq 1$$  where the infimum is over all
stopping times with respect to $Y$, then exhibit stopping
rules that achieve these bounds in the large-threshold regime and
large-delay-large-threshold regime, respectively. For noisy observations, two
complementary asymptotically optimal stopping rules are proposed. One depends on a single
observation at some fixed time but its optimality is usually very asymptotic. The other performs a sequential minimum mean
square error (mmse) estimate of $X_t$ given $Y_t$, $t=0,1,\ldots$ and stops as
soon as this estimate reaches level $\ell$. As such, the second stopping time
needs many more observations, roughly $\ell/s$, but performs
significantly better in the non-asymptotic regime. 

In the particular case where $X$ doesn't drift, we characterize $\inf_\eta\ex|\eta-\tau|^p$
non-asymptotically for both the noisy and the delayed observation cases.

Section~\ref{sec:mainresult} contains the main results and Section~\ref{sec:pf} is devoted to the
proofs.
\section{Results}\label{sec:mainresult}
Consider the discrete-time process
\begin{align*}
X:&\quad X_0=0\qquad X_t=\sum_{i=1}^t V_i +s t \qquad t\geq 1\,,
\end{align*}
where $s\geq 0$ is some known constant, where $V_1,V_2,\ldots$
are $\text{i.i.d.}\sim{\cal{N}}(0,1)$ (zero mean unit variance
Gaussian random variables), and consider the
first-passage time
$$
\tau_\ell=\inf\{t\geq 0: X_t\geq \ell\}
$$
for some known fixed threshold level $\ell\geq 0$.

Given sequential observations of a process
$Y=\{Y_t\}_{t\geq 0}$ correlated to $X$, we consider the
optimization problem
\begin{align}\label{oppb}
\inf_\eta\ex|\eta-{\tau_\ell}|^p, \: p\geq 1,
\end{align}
where the infimum is over all  stopping times $\eta$ defined
with respect to the natural filtration
induced by $Y$.\footnote{We 
consider only non-randomized stopping times since this does not induce a loss
of optimality with respect to \eqref{oppb} (see, e.g., \cite[Chap. 8.5]{DeG}
where randomization is shown to be useless for general statistical decision problems).}  

The results, presented in the next two subsections, relate to the situations
where $Y$ is either a noisy version of $X$, or a delayed version of $X$.

Throughout the paper the following notational conventions are adopted. We use
$\eta$ to denote a function of $Y=Y_0^\infty$. When
$\eta$ has no argument, such as in \eqref{oppb}, we mean that $\eta$ is a stopping time with
respect to $Y$. Instead, if $\eta$ has an
argument, we mean that $\eta$ is a function of its
argument which need not be a stopping time with
respect to $Y$.  For
example, $\eta(Y_a^b)$, with $0\leq a\leq b\leq
\infty$, refers
to a function of observations $Y_a^b=Y_a,Y_{a+1},\ldots,Y_b$.

 Further, we frequently omit arguments of functions (or estimators) that appear in
expressions to be optimized. For instance, instead of $$\inf_{\eta(Y_a^b)}\ex
|\eta(Y_a^b)-\tau_\ell|^p\,,$$ 
we simply write
$$\inf_{\eta(Y_a^b)}\ex |\eta-\tau_\ell|^p$$ 
to denote an optimization over estimators of
$\tau_\ell$ that depend only on observations $Y_a^b$.

\subsection{Noisy observations}
Consider the observation process
\begin{align*}
Y:&\quad Y_0=0\qquad Y_t=X_t+\varepsilon \sum_{i=1}^t W_i \quad
\:\: t\geq 1\,,
\end{align*}
where  $W_1,W_2,\ldots$ are i.i.d.$\sim{\cal{N}}(0,1)$ and where 
$\varepsilon\geq 0$ is some known constant. The observation noises
$\{W_i\}$ are supposed to be independent of $\{V_i\}$. 

Note that if $\ell=0$ or if $\varepsilon=0$ (i.e., $X=Y$), \eqref{oppb} is equal to zero by
setting $\eta=0$ and $\eta=\tau_\ell$, respectively. 

Interestingly, when  $\ell>0$, $\varepsilon>0$, and $s=0$, it turns out that it is impossible to track
$\tau_\ell$, even having access to the entire observation process
$Y_0^\infty$:
\begin{thm}[Noisy observations, $s=0$, \cite{BT} Proposition $2.1$.ii.]\label{prop1} For
$s=0$, $\varepsilon>0$, $\ell>0$, and $p\geq 1/2$, we have\footnote{Recall that
$\eta(Y_0^\infty)$ denotes an arbitrary function of observations $Y_0^\infty$ which
need not be a stopping time, according to our notational convention of the
previous section.}  $$\ex
| \eta(Y_0^\infty)-\tau_\ell |^p=\infty$$ for any estimator $\eta\left(Y_{0}^{\infty}\right)$ of $\tau_\ell$.
\end{thm}

We now consider the case $\ell>0$, $\varepsilon>0$, and $s>0$. The next result
characterizes \eqref{oppb} in the limit $\ell\to \infty$ and provides two
asymptotically optimal stopping rules. One of these rules is non-sequential in the
sense that it depends on a single observation. 

The sequential stopping rule is defined as
\begin{align}\label{etastar2}
\etas\defeq \inf\{t\geq 0:\hat{X}_t\geq \ell \}\,,\end{align} 
where $\hat{X}_0\defeq 0$ and where
\begin{align}\label{mmse0}
 \hat{X}_t\defeq
\frac{1}{1+\varepsilon^2}Y_t+\frac{s\varepsilon^2}{1+\varepsilon^2}t\qquad t\geq 1
\end{align}
is the mmse estimator of $X_t$ given
observation~$Y_t$.

The non-sequential stopping rule is defined as follows. Let\footnote{$x_+$
denotes $\max\{0,x\}$ and $\lfloor x\rfloor$ denotes
the integer part of $x$.}
\begin{align}\label{etastar}
\etans\defeq t^\star+\left\lfloor\frac{(\ell-\hat{X}_{t^\star})_+}{s}\right\rfloor\,,
\end{align}
with
\begin{align}\label{eq:n}t^\star\defeq \lfloor\ell/s-(\ell/s)^q\rfloor \,,
\end{align}
for some arbitrary constant $q\in (1/2,1)$. Notice that $\etans$ is only a function
of observation $Y_{t^\star}$.

\begin{thm}[Noisy observations, $s>0$]\label{th:main}
Fix $0< \varepsilon<\infty$, $0<s<\infty$, and $p\geq 1$. 
Then, for $\eta=\etas$ or $\eta=\etans$
\begin{align}\label{abiba}\ex
|\eta-\tau_\ell|^p&=(1+o(1))\inf_{\eta'(Y_0^\infty)} \ex
|\eta'-\tau_\ell|^p\nonumber \\
&=(1+o(1))C_1(\ell,s,\varepsilon,p)
\end{align}
as $\ell\rightarrow \infty$, where 
$$C_1(\ell,s,\varepsilon,p)\defeq
\left(\frac{\ell\varepsilon^2}{s^{3}(1+\varepsilon^2)}\right)^{p/2}{\ex}\left|
N\right|^p\,,$$
and where $N\sim{\cal{N}}(0,1)$. 
\end{thm}
Since $$\ex
|\eta-\tau_\ell|^p\geq \inf_{\eta'} \ex
|\eta'-\tau_\ell|^p\geq \inf_{\tilde{\eta}(Y_0^\infty)} \ex
|\tilde{\eta}-\tau_\ell|^p\,,$$ the first equality in \eqref{abiba} says that both
stopping rules $\etas$ and $\etans$ do as well as the best non-causal
estimators of $\tau_\ell$ with access to the entire observation process
$Y$, asymptotically. Moreover, note that asymptotic optimality is universal over $p\geq 1$
for $\etas$ and universal over both $p$ and $\varepsilon$ for $\etans$---since
the former does not depend on $p$ and the latter depends neither on $p$ nor on
$\varepsilon$. For $p=1$, the optimality of
$\etas$ was established in \cite[Theorem 2.3]{BT}.

Since $\etans$ does not exploit the dependency between $X$ and $Y$
($\etans$ does not depend on $\varepsilon$), it may be expected that
$\etas$ performs significantly better that $\etans$ for moderate to
low values of $\ell$. \begin{figure}
\begin{center} \includegraphics[scale=.29]{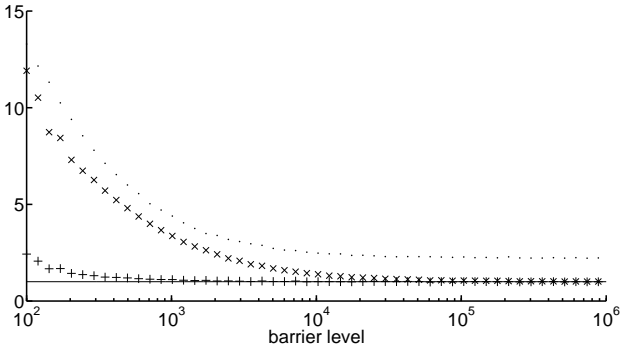} \caption{
$\ex|\eta-\tl|^p/C_1(\ell,s,\varepsilon,p)$ as a function of $\ell$
for $\eta=\etas$, $\eta=\etans$, and $\eta=\ell/s$ (marks 
$+$, $\times$, and $\bullet$, respectively), with $p=1$, $s=10$,
$\varepsilon=.5$, and $q=.51$. \label{fig}} \end{center} \end{figure}
In fact, this claim is supported numerically. An illustration is given by Fig.~\ref{fig} which represents numerical evaluations of 
\begin{align}\label{ler}
\frac{\ex|\eta-\tl|^p}{C_1(\ell,s,\varepsilon,p)} \end{align} as a
function of $\ell$ for $\eta\in\{\etas,\etans,\ell/s\}$, with parameters $p=1$,
$s=10$, and $\varepsilon=.5$. The parameter $q$ in the definition of
$\etans$ is chosen to be equal to $.51$. The simulation has a precision of
$\delta=.1$ for $\eta=\etas$ and $\eta=\etans$, and a precision of $\delta=.5$
for $\delta=\ell/s$. By precision we mean that the numerical evaluation of
\eqref{ler} deviates from it by less than $\delta$ with probability at least
$1-\delta$.
Simulation details are provided in the appendix.

We observe that, as $\ell\to \infty$,
\eqref{ler} tends to $1$ for both $\eta=\etas$ and $\eta=\etans$, as predicted by
Theorem~\ref{th:main}. However, $\etas$ performs significantly better than
$\etans$ in the non-asymptotic regime. For instance, for
$\ell\approx 1000$, $\ex
|\etas-\tau_\ell|$ is roughly a third of $\ex
|\etans-\tau_\ell|$.

More generally, simulation results suggest that 
$\ex|\etas-\tau_\ell|^p$ never exceeds $\ex
|\etans-\tau_\ell|^p$, and this for arbitrary $\ell>0$, $s>0$,
$\varepsilon>0$, and $p\geq 1$.\footnote{Parameter $q$ is kept equal
to $.51$ in our study.} Moreover, the difference between
$\ex|\etas-\tau_\ell|^p$ and $\ex
|\etans-\tau_\ell|^p$ increases as $\ell$ decreases, and can be very
significant for moderate to low
values of $\ell$. For instance, for $\ell=1000$, $s=10$, $\varepsilon=.1$,
and $q=.51$, we have $$(\ex|\etans-\tau_\ell|)/(\ex
|\etas-\tau_\ell|) \approx 12\:(!)$$
 
Thus, $\etans$ is suitable for very large values of $\ell$ since it has the interesting
feature of being a function of a single
observation. While also asymptotically optimal, $\etas$ does significantly better than $\etans$ in the non-asymptotic
regime, but requires roughly $\ell/s$ observations on average. To see this, note
that $\ex \hat{X}_\etas\approx
\ell$, and since $\hat{X}_t-\hat{X}_{t-1}=s$, we have $\ex \etas \approx \ell/s$
by Wald's equality---the approximations become equalities if we ignore
excess over the boundary (variously
known as ``overshoot''), i.e., that $\hat{X}_\etas$ may 
exceed $\ell$.
 
Concerning the fixed time estimator $\eta=\ell/s$, later it is shown (see
paragraph
after Lemma~\ref{lemma}) that 
\begin{align}\label{ft}
\lim_{\ell \to
\infty}\frac{\ex|\tl-\ell/s|^p}{C_1(\ell,s,\varepsilon,p)}=\left(\frac{1+\varepsilon^2}{\varepsilon^2}\right)^{p/2}
\end{align}
which is always greater than $1$. Hence $\eta=\ell/s$ is always suboptimal, and in
particular for small values of the noise parameter $\varepsilon$.
As $\varepsilon$ increases, the observation process $Y$ becomes
noisier and ultimately useless in the limit $\varepsilon\to \infty$. 
In this regime the fixed time estimator
$\ell/s$ is optimal. In the example of
Fig.~\ref{fig}, the right-hand side of \eqref{ft} is equal to $\sqrt{5}$.

\subsection{Delayed observations}
Consider the observation process\begin{align*}
Y:&\quad Y_0=0,Y_1=0,\ldots, Y_{d}=0\qquad Y_t=X_{t-d}
\quad t\geq d+1
\end{align*}
for some fixed positive integer $d\geq 0$.

Given $d\geq 0$, $\ell\geq 0$, and $s\geq 0$, define the stopping rule
$$\eta^{*}_d\defeq \inf\{t\geq 0:Y_t\geq \ell-s\cdot
d\}\,.$$
Notice that $\eta^{*}_d$ is a very natural candidate for estimating $\tau_\ell$
since, on average, $X_t$ is $s\cdot d$ higher than $Y_t$. In fact, the following
two theorems establish optimality of $\eta^{*}_d$ for any $s\geq 0$.
\begin{thm}[Delayed observations, $s=0$]\label{prop2} For
 $s=0$, $\ell>0$, and  $p\geq 1/2$, 
\begin{align*}
\inf_{\eta}{\ex}|\eta
-\tau_\ell|^p&=d^p={\ex}|\eta^{*}_d
-\tau_\ell|^p\,.
\end{align*}
\end{thm}
Instead, when the drift is positive we have: 
\begin{thm}[Delayed observations, $s>0$]\label{th:main2}
For $s>0$ and $p\geq 1$, 
\begin{align*}
\inf_{\eta}{\ex}|\eta
-\tau_\ell|^p&=(1+o(1)){\ex}|\eta^{*}_d
-\tau_\ell|^p\\
&= (1+o(1))C_2(d,s,p)
\end{align*}
as $d\rightarrow
\infty$ while $\ell=\ell(d)\geq s\cdot d$, where
$$C_2(d,s,p)\defeq \frac{d^{p/2}}{s^p}{\ex}|N|^p\,.$$
\end{thm}
In Theorem~\ref{th:main2}, note that $\ell$ need only be greater or equal than 
$s\cdot d$, and there is no other growth rate constraint of $\ell$ with respect
to $d$.

Also, notice that
$\eta^{*}_d$ is uniformly
optimal over $p\geq 1$, similarly as $\etas$ and $\etans$ for noisy observations.
However, by contrast with $\etas$ and $\etans$, optimality of $\eta^{*}_d$ is only with
respect to stopping times, not with respect to arbitrary functions
of $Y_0^\infty$. Indeed, if $\eta$ can be an arbitrary function of $Y_0^\infty$,
then we can set $\eta=\tl$ and so
achieve $\ex|\eta-\tau_\ell|^p=0$---in this case $\eta$ is no more a stopping time with respect to
$Y$ since
causality is violated.

Finally, note that for $s=0$ we have $\pr(\eta^{*}_d<\tau_\ell)=0$, i.e., it is optimal to wait until it is certain that $X$ reached
level $\ell$, and the corresponding estimation error is equal to $d^p$. By contrast, the estimation error grows as $d^{p/2}$ for
$s>0$. Thus, when $s>0$, were we to impose the additional certainty constraint
$\pr(\eta<\tau_\ell)=0$, the price to pay in terms of estimation error would be
a multiplicative factor of the order of $d^{p/2}$.

\begin{figure}
\begin{center}
\includegraphics[scale=.26]{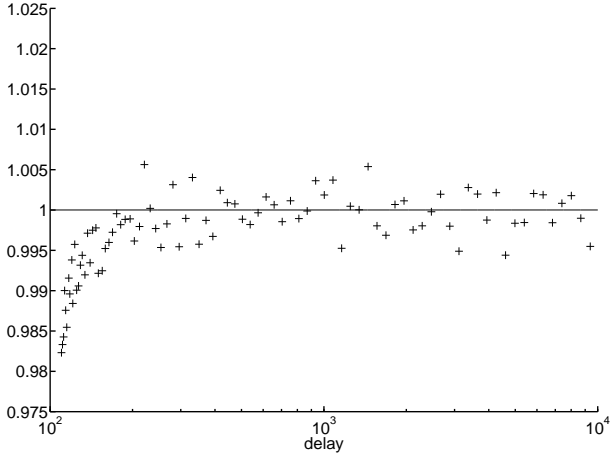}
\caption{ 
$\ex|\eta^{*}_d-\tl|^p/C_2(d,s,p)$ as a function of $d$ with
$\ell=100+s\cdot d$, $s=1$, $p=1$.
 \label{fig2}}
\end{center}
\end{figure}
Fig.~\ref{fig2} represents a numerical evaluation of 
\begin{align}
\frac{\ex|\eta^{*}_d-\tl|^p}{C_2(d,s,p)}\label{ller}\end{align} as a function of $d$ with
$\ell=100+s\cdot d$, for  $p=1$ and $s=1$. The function
is roughly equal to $1$, in agreement with Theorem~\ref{th:main2}. The small
oscillations around $1$ are due to our simulation which evaluates \eqref{ller}
 with a finite number of random samples. Here this number
suffices to guarantee a precision equal to
$\delta=.03$. Simulation details are provided in the
appendix. 

\subsection{Continuous time}
Theorems~\ref{prop1}, \ref{th:main}, \ref{prop2},
and \ref{th:main2} remain valid if we replace
$X$ and $Y$ by their continuous time
counterparts; i.e., $$X_t=s\cdot t+ B_t$$ and either
$$Y_t=X_t+\varepsilon W_t$$ for noisy observations,
or $$Y_t=X_{t-d}$$ for delayed observations, where
$$\{B_t\}_{t\geq 0}\qquad \text{and}\qquad \{W_t\}_{t\geq 0}$$
are independent standard Wiener processes. The proofs of the results in
continuous time are omitted since the arguments closely follow those in discrete time and often get
simplified as there is no issues
related to barrier overshoot.

\section{Proofs}\label{sec:pf}
In this section we prove first Theorems~\ref{th:main} and \ref{th:main2}, then
Theorem~\ref{prop2}. To prove Theorems~\ref{th:main} and \ref{th:main2}, we  often use the
following  Lemma, whose proof is deferred to the end of this section, on the
concentration of $\tau_\ell$ around its mean:
\begin{lem}\label{lemma} Let $S_t=\sum_{i=1}^t Z_i$ where $Z_1,Z_2,\ldots$ are i.i.d.
Gaussian random variables with mean 
$0<s<\infty$ and variance $0<\sigma^2<\infty$. Let $0<\ell<\infty$ and let
$$\mu=\inf\{t\geq 1: S_t\geq \ell\}\,.$$
Then, 
\begin{itemize}
\item[i.]
the following inequalities hold
\begin{equation}\label{stat2d}
\pr\left(\mu < \ell/s-z\right) \leq
\exp\left\{-\frac{s^2z^{2}}{2\sigma^2(\ell/s-z)}\right\}
\end{equation}
for $0 \leq z < \ell/s$;
\begin{equation}\label{stat2a}
\pr\left(\mu > \ell/s+z\right) \leq
\exp\left\{-\frac{s^2z^{2}}{2\sigma^2(\ell/s+z)}\right\} 
\end{equation}
for $z \geq 0$;
\item[ii.]
for any $p \geq 0$
\begin{align}
\label{stat2c}
\ex\left|\mu - \frac{\ell}{s}\right|^{p} \leq
k_1(k_2+\ell)^{p/2}
\end{align}
where $0\leq k_1,k_2<\infty$ are constants that depend on $p,s,\sigma^2$ but not on
$\ell$;
\item[iii.]  as $\ell \to \infty$, 
$$\sqrt{\frac{s^3}{\sigma^2\ell}}\left(\tau_\ell-\frac{\ell}{s}\right)\to
{\cal{N}}(0,1)$$
in distribution.
\end{itemize}
\end{lem}
Claim iii. of Lemma~\ref{lemma} implies \eqref{ft}. To see this, let $\tl$ be the
first time process $X$ reaches level $\ell$. Claim iii. of
Lemma~\ref{lemma} then  gives
\begin{align}\label{etals}\ex|\tl-\ell/s|^p=(1+o(1))\frac{\ell^{p/2}}{s^{3p/2}}\ex|N|^p\quad (\ell \to
\infty)
\end{align}
where $N\sim {\cal{N}}(0,1)$. This establishes \eqref{ft}. 

The following basic fact is repeatedly used in the proofs of Theorems~\ref{th:main}
and~\ref{th:main2}:
\begin{fact}
\label{fact}
Let $(S,Q)$ be two arbitrary random variables.  Then, 
$$\inf_{\eta(S)}\ex  |\eta\cdot
f(S)-g(S)-h(Q)|^p=\inf_{\eta(S)}\ex  |\eta
-h(Q)|^p\qquad p\geq 0$$
for any functions $g(\cdot)$ and $h(\cdot)$, and any function
$f(\cdot)$ such that $f(S)>0$ almost surely.
\end{fact}
To see this, notice first the obvious inequality
$$\inf_{\eta(S)}\ex  |\eta
f(S)-g(S)-h(Q)|^p\geq \inf_{\eta(S)}\ex  |\eta
-h(Q)|^p\,.$$
To see that
$$\inf_{\eta(S)}\ex  |\eta
f(S)-g(S)-h(Q)|^p\leq \inf_{\eta(S)}\ex  |\eta
-h(Q)|^p\,,$$
observe that for any $\eta=\eta(S)$ one can find
$\tilde{\eta}=\tilde{\eta}(S)$ such that $$\tilde{\eta}
f(S)-g(S)=\eta$$ almost surely since $f(S)>0$ almost surely.

To illustrate Fact~\ref{fact}, consider the
following simple example, variations of which 
appear in the proofs of Theorems~\ref{th:main} and \ref{th:main2}.

Let $X=Y+Z$ where $X$ and $Y$ are arbitrary random variables. Then, for any $c>0$
\begin{align*}
\inf_{\eta(Y)}\ex  |\eta
-c\cdot X|^p&=c^p\inf_{\eta(Y)}\ex  |\eta/c
-X|^p\nonumber \\
&=c^p\inf_{\eta(Y)}\ex  |\eta/c
-Y-Z|^p\nonumber \\
&=c^p
\inf_{\eta(Y)}\ex  |\eta
-Z|^p\,,
\end{align*}
where the last equality follows from Fact~\ref{fact} with $S=Y$, $Q=Z$, $f(S)=1/c$,
$g(S)=S $, and $h(Q)=Q$. 

We now prove Theorems~\ref{th:main} and \ref{th:main2}, then Theorem \ref{prop2}.
Throughout the proofs, $N$ always denotes a zero mean unit variance
Gaussian random variable.

\subsection{Proof of Theorem~\ref{th:main}}
We first show that
\begin{align}\label{oppb2}
\inf_{\eta(Y_0^\infty)}\ex |\eta-\tau_\ell|^p\geq (1+o(1))C_1(\ell,s,\varepsilon,p)\,,
\end{align}
where $C_1(\ell,s,\varepsilon,p)$ is defined in Theorem~\ref{th:main}, then show
that
$\ex|\eta-\tl|^p$ is equal to the right-hand side of \eqref{oppb2}  for 
$\eta=\etas$ and $\eta=\etans$. 
Before proceeding formally, we outline the main arguments.

To show \eqref{oppb2}, 
 the main idea is to reduce the minimization problem of estimating $\tl$ to the
 one of 
 estimating process $X$ at an instant
  close to $\ell/s$, the expected time
$X$ reaches level $\ell$. 
 To do this reduction, let $t^\star$ be such that ${t^\star}\approx \ell/s$ while satisfying $\pr(\tau_\ell\geq
 {t^\star})\approx 1$---one such instant is the $t^\star$ defined in
 \eqref{eq:n}.  It then follows that
 \begin{align}\label{tau}
 \tau_\ell \overset{\text{d}}{\simeq}
 t^\star+\frac{(\ell-X_{t^\star})_+}{s}\,,
 \end{align}
  since the time it takes
 for $X$ to go up by $q\geq 0$ is $q/s$ plus some
 small Gaussian term, by Claim iii. of
Lemma~\ref{lemma}. From \eqref{tau}, the fact that
$Y_{t^\star}$ is a
sufficient statistic for $X_{t^\star}$, and that $t^\star$ is close to $\ell/s$,
one can show
that \begin{align}
\inf_{\eta(Y_0^\infty)}
\ex|\eta-\tau_\ell|^p&\geq (1+o(1))\frac{1}{s^p}\inf_{\eta(Y_{t^\star})}\ex|\eta
-X_{t^\star}|^p 
\label{exlower}
\end{align}
where the infimum is
over estimators that
depend only on $Y_{t^\star}$.

Since $(X_{t^\star},Y_{t^\star})$ are jointly
Gaussian, for all $p\geq 1$ the infimum on the
right-hand side of \eqref{exlower} is
achieved by $\hat{X}_{t^\star}$, the mmse estimator \eqref{mmse0} of
$X_{t^\star}$ given observation $Y_{t^\star}$.  It then follows that
\begin{align*}
\inf_{\eta(Y_{t^\star})}\ex|\eta
-X_{t^\star}|^p=\left(\frac{\ell
\epsilon^2}{s(1+\varepsilon^2)}\right)^{p/2}\ex|
N|^p
\end{align*}
which, together with \eqref{exlower}, gives \eqref{oppb2}.

To achieve the right-hand side of \eqref{oppb2}, it is natural to consider the
stopping time
\begin{align}
\label{buu}
\etans=
t^\star+\left\lfloor\frac{(\ell-\hat{X}_{t^\star})_+}{s}\right\rfloor
\end{align}
which is similar to the right-hand side expression of \eqref{tau}, except that
$X_{t^\star}$ is
replaced by its (optimal) mmse estimator $\hat{X}_t$ (the discrepancy due to the
rounding in \eqref{buu} plays no role
asymptotically).

This stopping time is in fact optimal since the moments of
$\etans-\tau_\ell$ coincide with the right-hand side of
\eqref{oppb2}, asymptotically.  Finally, since $\hat{X}_t$ is the best estimator
of $X_t$, $\etas$ also represents a natural candidate since it is based on sequentially
estimating $X$ in an optimal fashion.

We proceed with the formal proof. 

 \noindent{{{\textbf{Lower bound:}}}} 
Fix $p\geq 1$ and fix an integer $t\geq 1$---later we take $t=t^\star$ defined
in \eqref{eq:n}. 

Then,
\begin{align}
\label{low31}& (\inf_{\eta\left(Y_0^{\infty}\right)}
{\ex}|\eta
-\tau_\ell|^p)^{1/p} \nonumber \\
&= \left(\inf_{\eta\left(Y_0^{\infty}\right)}{\ex}
\left|\left(\eta - t - \frac{\ell-X_{t}}{s}\right) -
\left(\tau_\ell-t- \frac{\ell-X_{t}}{s}\right)\right|^p\right)^{1/p}\nonumber \\
& \geq \left(\inf_{\eta\left(Y_0^{\infty}\right)}{\ex}
\left|\eta - t -
\frac{\ell-X_{t}}{s}\right|^p\right)^{1/p}\notag\\
&\hspace{3cm} -
\left({\ex}\left|\tau_\ell-t- \frac{\ell-X_{t}}{s} 
\right|^p\right)^{1/p} \nonumber \\
& =\left(\inf_{\eta(Y_t)}{\ex}
\left|\eta - t -
\frac{\ell-X_{t}}{s}\right|^p\right)^{1/p} \notag \\
&\hspace{3cm}-
\left({\ex}\left|\tau_\ell-t- \frac{\ell-X_{t}}{s} 
\right|^p\right)^{1/p} \,,
\end{align}
where the inequality holds by the triangle
inequality, and where the last equality holds since $Y_t$ is a
sufficient statistics for $X_t$. 

Since $(X_t,Y_t)$ are jointly Gaussian, 
\begin{align}\label{jgr}X_t\overset{\text{d}}{=}\hat{X}_t+\left(\frac{t\varepsilon^2
}{1+\varepsilon^2}\right)^{1/2}N\,,
\end{align}
where
$\hat{X}_t$ is the mmse estimator of $X_t$ 
given observation $Y_t$ defined in \eqref{mmse0}, and where
$N\sim{\cal{N}}(0,1)$ is independent of $\hat{X}_t$.

Hence,
\begin{align}\label{low32}
\inf_{\eta\left(Y_t\right)}{\ex}
\left|\eta - t -
\frac{\ell-X_{t}}{s}\right|^p &=\frac{1}{s^p}\inf_{\eta\left(Y_t\right)}{\ex}
\left|\eta s -ts-\ell- X_t
\right|^p\nonumber\\
&=\frac{1}{s^p}\inf_{\eta\left(Y_t\right)}{\ex}
\left|\eta - X_t
\right|^p\nonumber\\
&=\frac{1}{s^p}{\ex}
\left|\hat{X}_t - X_t
\right|^p\nonumber\\
&=\left(\frac{t \varepsilon^2
}{s^2(1+\varepsilon^2)}\right)^{p/2}{\ex}|N|^p\,.
\end{align}
The second equality follows from 
Fact~\ref{fact}.  The third equality holds since the mmse estimator of $X_t$
minimizes the average of any absolute moment with respect to $X_t$. The fourth
equality holds by \eqref{jgr}.

We now upperbound the second term on the right-hand side of~\eqref{low31}.
As we shall see, compared to the first term, the
contribution of the second term is negligible when $t=t^\star$.  

We have
\begin{align}\label{kigz}
&\ex|(\tau_\ell-t)-(\ell-X_{t})/s|^p\nonumber\\
&=
\ex(|(\tau_\ell-t)-(\ell-X_{t})/s|^p;\tau_\ell\leq
t)\notag \\
&\hspace{1.8cm}+\ex(|(\tau_\ell-t)-(\ell-X_{t})/s|^p;\tau_\ell>t)\,.
\end{align}

For the first term on the right-hand side of \eqref{kigz},
\begin{align}\label{kigz2}
\ex(|(\tau_\ell-t&)-(\ell-X_{t})/s|^p;\tau_\ell\leq t)\notag \\
&\leq
\ex\left((t+\ell/s+|X_{t}|/s)^p;\tau_\ell\leq t\right)\nonumber\\
&\leq \left[\ex (t+\ell/s+|X_{t}|/s)^{2p}\pr(\tau_\ell\leq t)\right]^{1/2}
\end{align}
by the triangle inequality and Cauchy-Schwartz inequality, respectively.

For the second term on the right-hand side of \eqref{kigz},
\begin{align}\label{cav1}
\ex&(|(\tau_\ell-t)-(\ell-X_{t})/s|^p;\tau_\ell>t)\nonumber\\
&=\ex\left(|(\tau_\ell-t)-(\ell-X_{t})/s|^p\big|\tau_\ell>t\right)\pr(\tau_\ell>t)\nonumber \\
&\leq \ex\left(\ex\left(|(\tau_\ell-t)-(\ell-X_{t})/s|^p\Big|X_{t},
\tau_\ell>t\right)\Big|\tau_\ell>t\right)\nonumber\\
&\leq  k_1\ex\left(k_2+(\ell-X_t)_+)^{p/2}\right)
\end{align}
where the second inequality follows from Claim ii. of Lemma~\ref{lemma}
and the strong Markov property of 
$X$ at time $t$, with $k_1,k_2\geq 0$ being constants that
depend only on $p$ and $s$.

Combining \eqref{low31}, ~\eqref{low32}, \eqref{kigz},
\eqref{kigz2}, and \eqref{cav1} yields
\begin{align}\label{izico}
(\inf_{\eta(Y_0^\infty)}{\ex}|\eta -\tau_\ell&|^p)^{1/p}\geq
\left(\frac{1}{s^p}\left(\frac{t\varepsilon^2}{(1+\varepsilon^2)}\right)^{p/2}{\ex}\left|
N\right|^p\right)^{1/p}\nonumber\\
&-\Big(
\big[\ex (t+\ell/s+|X_{t}|/s)^{2p}\pr(\tau_\ell\leq t)\big]^{1/2}\notag \\
&+k_1\ex\left(k_2+(\ell-X_t)_+\right)^{p/2}\Big)^{1/p}.
\end{align}
Finally, letting $t=t^\star$ where $t^\star$ is defined in \eqref{eq:n},
we have
$$
\pr (\tau_\ell\leq t^\star)\leq \exp(-\Omega(\ell^{2q-1}))
$$ by Claim~$i.$ of
Lemma~\ref{lemma}.\footnote{$\Omega(\cdot)$ refers
to standard order notations, see, e.g.,
\cite[Chapter 3]{CLRS}.} Therefore,
\begin{align}\label{yiu}
\ex (t^\star+\ell/s+|X_{t^\star}|/s)^{2p}\pr(\tau_\ell\leq
t^\star)=o(1)\qquad (\ell\to
\infty)
\end{align}
since 
\begin{align}
\label{grepre}
X_{t^\star}\overset{\text{d}}{=}s\cdot
t^\star+(t^\star)^{1/2}N\,.
\end{align}
 From \eqref{grepre} and \eqref{eq:n} we also get 
\begin{align}\label{yiu2}
\ex\left(k_2+(\ell-X_{t^\star})_+\right)^{p/2}&=O(\ell^{q p/2})\nonumber \\
&=o(\ell^{p/2})
\end{align}
since $q<1$.
From \eqref{izico} with $t=t^\star$,  \eqref{yiu}, and \eqref{yiu2} we get
\begin{align}\label{loublie}
\inf_{\eta(Y_0^\infty)}{\ex}|\eta -\tau_\ell|^p\geq
(1+o(1))\left(\frac{\ell\varepsilon^2}{s^{3}(1+\varepsilon^2)}\right)^{p/2}{\ex}\left|
N\right|^p
\end{align}
as $\ell\to \infty$, yielding the desired result.

Next, we establish the asymptotic optimality of $\etas$ and
$\etans$ by showing that their absolute moments with respect to
$\tl$ is equal to the right-hand side
of \eqref{loublie}. The proof of optimality of $\etas$
uses most of the arguments of the proofs of \cite[Theorem $2.1$]{BT},
which establishes optimality of $\etas$ for $p=1$, together with
some of the arguments used to establish optimality of $\etans$. 

\noindent{{\textbf{Achievability, $\etans$:}}} 
 To simplify exposition, we ignore
discrepancies due to the rounding of non-integer quantities as they play no role
asymptotically. In particular, we assume that $\etans$ is given
by
$$\etans=t^\star+\frac{(\ell-\hat{X}_{t^\star})_+}{s}$$
without rounding the fraction.\footnote{As such, $\etans$ is no more
a stopping time, strictly speaking.} Notice that if $\etans$,
as defined above, is asymptotically optimal, then a
triangle inequality argument immediately shows that
$\etans$ with the rounding of the fraction is also
asymptotically optimal.

Let 
\begin{align}
\label{del}
\Delta\defeq\tau_\ell - t^\star\,,
\end{align}
 and let 
\begin{align}\label{delhat}
\hat{\Delta}\defeq   (\ell-\hat{X}_{t^\star})_+/s\,.
\end{align}
Then,
 \begin{align}\label{eq:1}
 \ex|\etans-\tau_\ell|^p&=
\ex
(|\hat{\Delta}-\Delta|^p;{\tau_\ell>{t^\star}})\notag \\
&\hspace{.5cm}+
 \ex(|\etans-\tau_\ell|^p ;\tau_\ell\leq {t^\star}) \,.
 \end{align}
  For the first term on the right-hand side of \eqref{eq:1}, 
 \begin{align}\label{eq:2}
\ex(|\hat{\Delta}-\Delta|^p;\tau_\ell>{t^\star})\leq\ex(|\hat{\Delta}&-\Delta|^p;\hat{X}_{t^\star}<\ell,\tau_\ell>{t^\star})\nonumber
\\
&+\ex(|\Delta|^p;\hat{X}_{t^\star}\geq
\ell)\,.
\end{align}
By the triangle inequality,
\begin{align}\label{eq:z}
(\ex(&|\hat{\Delta}-\Delta|^p;\hat{X}_{t^\star}<\ell,\tau_\ell>t^\star))^{1/p}\nonumber\\
&\leq(\ex(|\hat{\Delta}-(\ell-X_{{t^\star}})/s|^p;\hat{X}_{t^\star}<\ell,\tau_\ell>{t^\star}))^{1/p}\nonumber\\
&+(\ex(|(\ell-X_{t^\star})/s-\Delta|^p;\hat{X}_{t^\star}<\ell,\tau_\ell>t^\star))^{1/p}\nonumber\\
&\leq(\ex(|\hat{\Delta}-(\ell-X_{t^\star})/s|^p;\hat{X}_{t^\star}<\ell))^{1/p}\nonumber\\
&+(\ex(|(\ell-X_{t^\star})/s-\Delta|^p;\tau_\ell>{t^\star}))^{1/p}\,.
\end{align}
For the first term on the right-hand side of \eqref{eq:z}, 
\begin{align}\label{eq:7}
\ex(|\hat{\Delta}-(\ell-X_{t^\star})/s&|^p;\hat{X}_{t^\star}<\ell)\notag\\
&=\ex(|(X_{t^\star}-\hat{X}_{t^\star})/s|^p;\hat{X}_{t^\star}<\ell)\nonumber\\
&\leq \ex|(X_{t^\star}-\hat{X}_{t^\star})/s|^p\nonumber \\
&=\frac{1}{s^p}\left(\frac{t^\star\varepsilon^2}{1+\varepsilon^2}\right)^{p/2}\ex|N|^p
\end{align}
where the last equality follows from \eqref{jgr}.

 For the second term on the right-hand side of  
\eqref{eq:z} we use \eqref{cav1} with $t=t^\star$ to get
\begin{align}\label{eq:5}
\ex(|(\ell-X_{t^\star})/s-\Delta|^p;\tau_\ell>t^\star)&\leq
k_1\ex\left(k_2+(\ell-X_{t^\star})_+\right)^{p/2}
\end{align}
where $k_1,k_2$ are constants that depend on $p$ and $s$ only.
 
 For the second term on the right-hand side of
\eqref{eq:2},  Cauchy-Schwartz inequality yields
 \begin{align}\label{eq:9}
 \ex(|\Delta|^p;\hat{X}_{t^\star}\geq
 \ell)\leq (\ex(|\Delta|^{2p}))^{1/2}\pr(\hat{X}_{t^\star}\geq
 \ell)^{1/2}\,.
\end{align}
By the triangle inequality,
\begin{align}\label{eq99}
(\ex|\Delta|^{2p})^{1/2p}&\leq
(\ex|\tau_\ell-\ell/s|^{2p})^{1/2p}+(\ex|\ell/s-t^\star|^{2p})^{1/2p}\nonumber \\
&
\leq k_1(k_2+\ell)^{1/2}+(\ell/s)^q\,,
\end{align}
where for the second inequality we used Claim ii. of
Lemma~\ref{lemma}, with $k_1,k_2$ constants that depend on
$p$ and $s$, and the
definition of $t^\star$ (recall that we ignore 
discrepancies due to the rounding of non-integer quantities).

From \eqref{eq:2}, \eqref{eq:z}, \eqref{eq:7},
\eqref{eq:5},  \eqref{eq:9}, and \eqref{eq99} we obtain
\begin{align}\label{eq:10}
\ex(|\hat{\Delta}-\Delta|^p;\tau_\ell>t^\star)&\leq
\Bigg[\left(\frac{1}{s^p}\left(\frac{t^\star\varepsilon^2}{(1+\varepsilon^2)}\right)^{p/2}\ex|N|^p\right)^{1/p}\nonumber\\
&\hspace{-2cm}+\left(k_1\ex\left(k_2+(\ell-X_{t^\star})_+)^{p/2}\right)\right)^{1/p}\bigg]^{p}\nonumber\\
&\hspace{-2cm}+ [k_1(k_2+\ell)^{1/2}+(\ell/s)^q]^p\pr(\hat{X}_{t^\star}\geq \ell)^{1/2}\,.
\end{align}
For the second term on the
right-hand side of \eqref{eq:1},  using  Cauchy-Schwartz
inequality and
the triangle inequality we get
\begin{align}\label{eq:11}
&\ex(|\tau_\ell-\etans|^p;\tau_\ell\leq t^\star)\notag \\
&\leq
\left[(\ex|\tau_\ell-\ell/s|^{2p})^{1/2p}+(\ex|\ell/s-\etans|^{2p})^{1/2p}\right]^{p}\pr(\tau_\ell\leq
t^\star)^{1/2}\nonumber\\
&\leq\left[ k_1 (k_2+ \ell)^{1/2}
+k_3 (k_4+ \ell)^{1/2}\right]^{p}\pr(\tau_\ell\leq t^\star)^{1/2}
\end{align}
where for the second inequality we used Claim ii. of
Lemma~\ref{lemma}, with ${k_3}, k_4$  constants that depend
on $p$, $s$, and $\varepsilon$.

Combining \eqref{eq:1}, \eqref{eq:10}, and \eqref{eq:11}
\begin{align}\label{eq:fin1}
\ex|\etans-\tau_\ell|^p&\leq
\left[\left(\frac{1}{s^p}\left(\frac{t^\star\varepsilon^2}{(1+\varepsilon^2)}\right)^{p/2}\ex|N|^p\right)^{1/p}\right.\nonumber\\
&+\left(k_1\ex\left(k_2+(\ell-X_{t^\star})_+)^{p/2}\right)\right)^{1/p}\Bigg]^p\nonumber\\
&+[k_1(k_2+\ell)^{1/2}+(\ell/s)^q]^p\pr(\hat{X}_{t^\star}\geq \ell)^{1/2}
\nonumber \\
&+ \left[ k_1 (k_2+ \ell)^{1/2}
+k_3
(k_4+\ell)^{1/2}\right]^{p}\pr(\tau_\ell\leq t^\star)^{1/2}\,.
\end{align}
Using \eqref{mmse0} and Claim~i. of Lemma~\ref{lemma}
one deduces that the
third and fourth terms on the right-hand side of
\eqref{eq:fin1} tend to zero as $\ell\to\infty$.
Since $X_{t^\star}\overset{\text{d}}{=}s\cdot
t^\star+(t^\star)^{1/2}N$ and $t^\star= (\ell/s)(1+o(1))$, we conclude
that
 \begin{align*}
\ex|\etans-\tau_\ell|^p&\leq(1+o(1))
\left(\frac{\ell\varepsilon^2}{s^3(1+\varepsilon^2)}\right)^{p/2}\ex|N|^p\\
&=(1+o(1))C_1(\ell,s,\varepsilon,p)
\end{align*}
as $\ell\to \infty$, 
where 
$$C_1(\ell,s,\varepsilon,p)\defeq
\left(\frac{\ell\varepsilon^2}{s^{3}(1+\varepsilon^2)}\right)^{p/2}{\ex}\left|
N\right|^p\,.$$
This establishes the asymptotic optimality of $\etans$.

\noindent{{\textbf{Achievability, $\etas$:}}} 
We write $\ex |\etas -\tl|^p$ as
\begin{align}\label{mwz}
\ex |\etas -\tl|^p&=\ex (|\etas -\tl|^p;\etas\geq \tl)\nonumber \\
&\hspace{2cm}+\ex
(|\tl-\etas|^p;\tl\geq \etas)\,,
\end{align}
and upper bound each of the two terms on right-hand side of the above equation. As in the previous section, we ignore
discrepancies due to the rounding of non-integer quantities as they play no role
asymptotically. In particular, we treat $\ell/s$ as an integer.

Letting
$$\nu\defeq \inf\{t\geq 0:\hat{X}_{\tl+t}\geq \ell\}\,,$$
we have
\begin{align}\label{m1}
\ex (&|\etas -\tl|^p;\etas\geq \tl)\nonumber \\
&=\ex (\nu^p;\etas\geq \tl)\nonumber \\
& \leq \ex (\nu^p;\hat{X}_\tl< \ell)\nonumber \\
& \leq \bigg[[\ex (\ell-\hat{X}_\tl)^p;\hat{X}_\tl<\ell)]^{1/p}/s\nonumber \\
&\hspace{.5cm}+[\ex |\nu-(\ell-\hat{X}_\tl)/s|^p;\hat{X}_\tl<
\ell)]^{1/p}\bigg]^{p}\nonumber\\
& \leq \bigg[[\ex (X_\tl-\hat{X}_\tl)_+^p]^{1/p}/s\nonumber \\
&\hspace{.5cm}+[\ex |\nu-(\ell-\hat{X}_\tl)/s|^p;\hat{X}_\tl<
\ell)]^{1/p}\bigg]^{p}\,,
\end{align}
where the first inequality follows from the definition of $\hat{X}_t$ (see
\eqref{etastar2}) and where the second inequality follows from the triangle
inequality.

 We upper bound
the two expectations on the right-hand side of \eqref{m1}.

For the first term, for $i\geq 1$ let 
\begin{align}\label{ui}
U_i&\defeq (X_i-\hat{X}_i)-(X_{i-1}-\hat{X}_{i-1})\nonumber\\
&=(\eps^2/(1+\eps^2))V_{i}-(\eps/(1+\eps^2)) W_{i}\\
&\overset{\text{d}}{=}\frac{\varepsilon}{(1+\varepsilon^2)^{1/2}}N\label{jmom}\,.
\end{align}
Then,\footnote{$\openone\{{\cal{A}}\}$ denotes the indicator function of event
$\cal{A}$.}
\begin{align}
{X}_{\tl}-\hat{X}_{\tl} 
=\sum_{i=1}^{\ell/s}U_{i} &-
\openone\{\tl<\ell/s\}\sum_{i=\tau_\ell+1}^{\ell/s}U_{i}\notag \\
&+
\openone\{\tl>m\} \sum_{i=\ell/s+1}^{\tl}U_{i} \,,
\end{align}
and, by the triangle inequality,\footnote{By $x_+^p$ we actually mean
$(x_+)^p$.}
\begin{align}\label{sumpo}
[\ex( {X}_{\tl}- \hat{X}_{\tl})_+^p]^{1/p} &\leq
\left[\ex\Big(\sum_{i=1}^{\ell/s}U_{i}\Big)_+^p\right]^{1/p}\notag \\
&+ \left[\ex\Big( -
\openone\{\tl<\ell/s\}\sum_{i=\tl+1}^{\ell/s}U_{i}\Big)_+^p\right]^{1/p}\nonumber \\
&+\left[\ex\Big( \openone\{\tl>\ell/s\}
\sum_{i=\ell/s+1}^{\tl}U_{i}\Big)_+^p\right]^{1/p}.
\end{align}
We bound each term on the right-side of \eqref{sumpo}. For the
first term, from \eqref{jmom} we have 
\begin{align}\label{ugo3}
{\ex}\left(\sum_{i=1}^{\ell/s}U_{i}\right)_{+}^p &= ((\ell/s)\eps^2/(1+\eps^2))^{p/2}\ex
N_+^p\,.
\end{align}
For the second term on the right-side of \eqref{sumpo}, using \eqref{jmom}
together with the fact that $\tl$ is independent of $U_{\tl+1},U_{\tl+2},\ldots
$ we get
\begin{align}\label{dzp}
\ex\Big( - \openone\{\tl<\ell/s\}&\sum_{i=\tl+1}^{\ell/s}U_{i}\Big)_+^p\notag
\\
&=
{\ex}[(\ell/s-\tl)_+\eps^2/(1+\eps^2)]^{p/2}\ex\,N_+^p \nonumber \\
& \leq  (\ex |\ell/s-\tl |^{p/2}){\ex}N_{+}^p\nonumber \\
&\leq k_1(k_2+\ell)^{p/4} {\ex}N_{+}^p  \nonumber \\
&=O(\ell^{p/4})\,,
\end{align}
where for the first inequality we bounded $\eps^2/(1+\eps^2)$ by $1$, and where
for the second inequality we used Claim ii. of Lemma~\ref{lemma}.

For the third term on the right-side of \eqref{sumpo}, using \eqref{ui}, the
triangle inequality, and by upperbounding $\varepsilon^2/(1+\varepsilon^2)$ and
$\varepsilon/(1+\varepsilon^2)$ by $1$, we get
\begin{align}\label{gk}
\Big(\ex\big(& \openone\{\tl>\ell/s\}
\sum_{i=\ell/s+1}^{\tl}U_{i}\big)_+^p\Big)^{1/p} \nonumber \\
&\leq
 {\ex}
\Big(\big(\openone\{\tl>\ell/s\}\sum_{i=\ell/s+1}^{\tau}W_{i}\big)_{+}^p\Big)^{1/p}\nonumber \\
&+\Big({\ex}
\big(\openone\{\tl>\ell/s\}\sum_{i=\ell/s+1}^{\tau}V_{i}\big)_{+}^p\Big)^{1/p}.
\end{align}
Since $\tl$ and $\{W_{i}\}$ are independent, we have
$$\openone\{\tl>\ell/s\}\sum_{i=\ell/s+1}^{\tl}W_{i} \overset{\text{d}}{=}
\sqrt{(\tl-\ell/s)_{+}}N\,,$$ and a similar calculation as
for~\eqref{dzp} shows that
\begin{align}\label{gk1}
{\ex} \big(\openone\{\tl>\ell/s\}\sum_{i=\ell/s+1}^{\tl}W_{i}\big)_{+}^p
=O(\ell^{p/4}).
\end{align}
We now focus on the second expectation on the right-side
of~\eqref{gk}. Since, on $\{\tl > \ell/s\}$, we have
$$
\sum_{i=\ell/s+1}^{\tl}V_{i} = (X_{\tl} - X_{\ell/s}) - s(\tl - \ell/s)\,,
$$
 we consider the shifted process $\{X_{t} - X_{\ell/s}\}_{t\geq
\ell/s}$ and its crossing of level $\ell - X_{\ell/s}$. It then follows that 
\begin{align}\label{gk3}
&{\ex}
\big(\openone\{\tl>\ell/s\}\sum_{i=\ell/s+1}^{\tl}V_{i}\big)_{+}^p\nonumber \\
&= s^p {\ex}\big(\left[(X_{\tl} - X_{\ell/s})/s - (\tl -
\ell/s)\right]_{+}^p;\tl>\ell/s, X_{\ell/s}<\ell\big)\nonumber \\
&\leq  s^p {\ex}\left(|(X_{\tl} - X_{\ell/s})/s - (\tl -
\ell/s)|^p\big |\tl>\ell/s,X_{\ell/s}<\ell\right)\nonumber \\
&\leq k_1{\ex} (k_2+(X_\tl-X_{\ell/s})_+)^{p/2}\nonumber \\
& =O(\ell^{p/4})\end{align}
where $k_1,k_2$ are constants that depend only on $s$ and $p$, and where the
second inequality follows Claim ii. of Lemma~\ref{lemma} and the Markov
property of process $X$ at time $\ell/s$. We now justify the second equality in
\eqref{gk3}. We have
$$X_{\ell/s}\overset{\text{d}}{=}\ell+\sqrt{\ell/s}N$$
and $$X_\tl=\ell+\e_\tl\,,$$ where $\e_\tl$ denotes the excess over the
boundary at time $\tl$.
Using this and the triangle inequality we get
\begin{align}
\left(\ex(X_\tl-X_{\ell/s})_+^{p/2}\right)^{2/p}\leq (\ex
\e_{\tl}^{p/2})^{2/p}+\sqrt{\ell/s }(\ex N_+^{p/2})^{2/p}\,,
\end{align}
which implies that
$$\ex(X_\tl-X_{\ell/s})_+^{p/2}=O(\ell^{p/4})$$
since $\ex
\e_{\tl}^{p/2}$ can be upper bounded by a finite constant that is 
independent of $\ell$ (\cite[Equation $(2)$]{Mog1}).
This establishes the second equality in \eqref{gk3}.

Combining \eqref{gk} together with \eqref{gk1} and \eqref{gk3}
yields
 \begin{align}\label{gk2}
\ex\Big( \openone\{\tau>{\ell/s}\}
\sum_{i={\ell/s}+1}^{\tau}U_{i}\Big)_+^p&=O(\ell^{p/4})\,.
\end{align}
From \eqref{sumpo}, \eqref{ugo3}, \eqref{dzp}, and
\eqref{gk2} we get
\begin{align}\label{zh}
\ex( {X}_\tl- \hat{X}_\tl)_+^p & \leq (1+o(1))\left( \frac{\ell
\eps^2}{s(1+\eps^2)}\right)^{p/2}\ex
N_+^p.
\end{align}

For the second expectation on the right-hand side of \eqref{m1} we have
\begin{align}\label{gk5}
\ex |\nu-(\ell-\hat{X}_\tl)/s|^p;\hat{X}_\tl< \ell)]&\leq k_3\ex
[k_4+(\ell-\hat{X}_\tl)_+]^{p/4}\nonumber \\
&=O(\ell^{p/4})\,,
\end{align}
where the inequality follows from the strong Markov property of $\hat{X}$
at time $\tl$ together with Claim ii. of Lemma~\ref{lemma}, with $k_3$ and $k_4$
constants that depend on $s$ and $\varepsilon$.

From \eqref{m1}, \eqref{zh}, and \eqref{gk5} we get
\begin{align}\label{ixi}
\ex (|\etas-\tl|^p;\etas \geq \tl) \leq (1+o(1))\left( \frac{\ell \eps^2}{s^3(1+\eps^2)}\right)^{p/2}\ex
N_+^p.
\end{align}
Using analogous arguments as for establishing \eqref{ixi}, which essentially
amounts to swap the roles of $X$ and $\hat{X}$ and the roles of $\tl$ and
$\eta_\ell^*$, we get
\begin{align}\label{ixi2}
\ex (|\tl-\etas|^p;\tl\geq \etas) \leq (1+o(1))\left( \frac{\ell \eps^2}{s^3(1+\eps^2)}\right)^{p/2}\ex
N_+^p.
\end{align}
Finally, from \eqref{mwz}, \eqref{ixi}, and \eqref{ixi2} we get
$$\ex |\tl-\etas|^p \leq (1+o(1))\left( \frac{\ell \eps^2}{s^3(1+\eps^2)}\right)^{p/2}\ex
|N|^p \qquad (\ell\to \infty)\,,$$
which establishes the asymptotic optimality of
$\etas$.\hfill{$\blacksquare$}

\subsection{Proof of Theorem~\ref{th:main2}} 
As mentioned earlier, $\eta^{*}_d$ is 
a very natural stopping time to consider since, on average, $X_t$ is $s\cdot d$
higher than $Y_t$. Now, the time needed to go from level $\ell-s\cdot d$ to level $\ell$
has (approximately) the Gaussian distribution $d+(\sqrt{d}/s) N$ by Claim iii.
of Lemma~\ref{lemma}. Hence we have
$\tau_\ell-\eta^{*}_d\overset{\text{d}}{\approx}
(\sqrt{d}/s) N$ which yields the second equality in Theorem~\ref{th:main2}. The
optimality of
$\eta^{*}_d$ is established essentially by showing that any (asymptotically) optimal stopping
rule shouldn't stop later than $\eta^{*}_d$.

 \noindent{{{\textbf{Lower bound:}}}} 
 Let $\ell$ be any function of $d$ such that $\ell\geq s\cdot d$, and fix 
 integer $d\geq 1$. Further, let  
 $$\nu\defeq \inf\{t\geq 0:X_t\geq \ell-s\cdot d (1-\varepsilon)\}$$  where $\varepsilon$ is a constant such that
$0<\varepsilon<1$---later we  take $\varepsilon \to 0$. 
 
Then,
 \begin{align}\label{eq:star}
 \inf_{\eta}{\ex}|\eta -\tau_\ell|^p&\geq 
\inf_{\eta}{\ex}(|\eta -\tau_\ell|^p;\tau_\ell\leq \nu+ d)\nonumber\\
& \geq \inf_{\eta(Y_0^{\nu+d})\leq \nu+d}{\ex}(|\eta
-\tau_\ell|^p;\tau_\ell\leq \nu+d)\nonumber\\
&=\inf_{\eta(X_0^{\nu})\leq \nu+d}{\ex}(|\eta
-\tau_\ell|^p;\tau_\ell\leq \nu+d)\,,
\end{align}
where the infimum on the right-hand side of the second
inequality
is over all estimators that depend on $Y_0^{\nu+d}$ (these estimators need
not be stopping times), and where  the equality holds since $Y_t=X_{t-d}$.

Let $$\delta\defeq \inf\{t\geq 0:X_{\nu+t}\geq \ell\}\,,$$
so that, by definition,  $$\tau_\ell=\nu+\delta\,.$$ 
Then,
\begin{align}\label{bmw}
\inf_{\eta(X_0^{\nu})\leq \nu+d}&{\ex}(|\eta -\tau_\ell|^p;
\tau_\ell\leq \nu+d)\notag \\
&=\inf_{\eta(X_0^{\nu})\leq \nu+d}{\ex}(|\eta
-(\nu+\delta)|^p;0\leq \delta \leq d)\nonumber\\
&=\inf_{\eta(X_0^{\nu})\leq \nu+d}{\ex}(|\eta
-\delta|^p;0\leq \delta \leq d)\nonumber\\
&=\inf_{ \eta(X_{\nu})\leq \nu+d}{\ex}(|\eta
-\delta|^p;0\leq \delta \leq d)\nonumber\\
&\geq \inf_{ \eta(X_{\nu})}{\ex}(|\eta
-\delta|^p;0\leq \delta \leq d)\nonumber\\
&\geq \inf_{ \eta(X_{\nu})}{\ex}(|\eta
-\delta|^p;0\leq \delta \leq d,\e_\nu\leq c)\,.
\end{align}
The second equality in \eqref{bmw} follows from
Fact~\ref{fact}.
The infimum on the right-hand side of the third
equality is over  estimators that depend on 
$X_{\nu}$ only, since $\delta$ is defined over $X_\nu,
X_{\nu+1},\ldots$. The last inequality holds for an arbitrary fixed constant
$c>0$, with  $\e_\nu$ defined as the excess at time
$\nu$, i.e.,
$$\e_\nu\defeq X_\nu-(\ell-s\cdot d(1-\varepsilon))\geq 0\,.$$

Take  $d$ large enough so that 
\begin{align}\label{cciw}
sd\varepsilon>c\,,
\end{align}
and define
\begin{align*}
d_\nu\defeq 
d(1-\varepsilon)+\e_\nu/s\,,
\end{align*}
\begin{align*}
N_\nu\defeq \sqrt{\frac{s^2}{d_\nu}}(\delta-d_\nu)\,,
\end{align*}
and define the functions $f_1(d,\varepsilon)$ and
$f_2(d,\varepsilon)$ as
$$f_1(d,\varepsilon)\defeq
s\sqrt{d(1-\varepsilon)}\,,$$
and 
$$f_2(d,\varepsilon)\defeq \frac{s
d\varepsilon-c}{\sqrt{d(1-\varepsilon)+c/s}}\,.$$
Notice that both $f_1$ and $f_2$ are strictly positive because of \eqref{cciw}.

 Using the definitions of $d_\nu$ and $N_\nu$ we get
\begin{align}\label{hgw}
 &\inf_{ \eta(X_{\nu})}{\ex}(|\eta
-\delta|^p;0\leq \delta \leq d,\e_\nu\leq c)\nonumber\\
&=\inf_{ \eta(X_{\nu})}{\ex}(|\eta
-(\delta-d_\nu)|^p;\E_1)\nonumber
\\&
\geq \inf_{ \eta(X_{\nu})}{\ex}\left(|\eta
-(\delta-d_\nu)|^p;\E_2\right)
\nonumber
\\&
\geq \frac{({d(1-\varepsilon)})^{p/2}}{s^p}\inf_{
\eta(X_{\nu})}{\ex}\left(|\eta\sqrt{s^2/d_\nu}
-N_\nu|^p;\E_2\right)
\nonumber \\
&= \frac{({d(1-\varepsilon)})^{p/2}}{s^p} \inf_{ \eta(X_{\nu})}{\ex}\left(|\eta
-N_\nu|^p;\E_2\right)\,.
\end{align}
where we defined the events
\begin{align*}
\E_1&\defeq \{-s\sqrt{d_\nu}\leq N_\nu\leq
s(d-d_\nu)/\sqrt{d_\nu},\e_\nu\leq c\}\\
\E_2&\defeq \{-f_1(d,\varepsilon)\leq
N_\nu\leq f_2(d,\varepsilon),\e_\nu\leq c\}\,.
\end{align*}
The first equality in \eqref{hgw} holds by Fact~\ref{fact}.
The first inequality holds by  the definitions of
$f_1(d,\varepsilon)$ and $f_2(d,\varepsilon)$ and by
noting that, on
$\{\e_\nu\leq c\}$, the range of $N_\nu$ in $\E_1$ contains the range of $N_\nu$
in $\E_2$. 
The second inequality holds by the definition of $N_\nu$
and because on event $\E_2$ we have 
\begin{align*}
d_\nu\geq d(1-\varepsilon)\,.
\end{align*}
 Finally the
last equality in \eqref{hgw} holds by Fact~\ref{fact} since $d_\nu$ is a
function of $X_\nu$ (through $e_\nu$).

Since $f_1(d,\varepsilon)$ and
$f_2(d,\varepsilon)$ are increasing functions of $d$, 
let us pick $d$  so that the following inequality, more stringent than
\eqref{cciw}, is satisfied 
\begin{align}\label{condc}
c< \min\{sd\varepsilon,f_1(d,\varepsilon),f_2(d,\varepsilon)\}.
\end{align}
It then follows that
\begin{align*}
{\ex}&\left(|\eta
-N_\nu|^p;-f_1(d,\varepsilon)\leq
N_\nu\leq f_2(d,\varepsilon),\e_\nu\leq c\right)\\
&\hspace{2cm}\geq  {\ex}\left(|\eta
-N_\nu|^p;-c\leq
N_\nu\leq c,\e_\nu\leq c\right)\,,
\end{align*} 
hence, from \eqref{hgw},
\begin{align}\label{boulgo}
&\frac{s^p}{({d(1-\varepsilon)})^{p/2}}\inf_{\eta(X_\nu)}{\ex}(|\eta
-\delta|^p;0\leq \delta \leq d,\e_\nu\leq c)\nonumber\\
&\geq \inf_{\eta(X_\nu)} {\ex}\left(|\eta
-N_\nu|^p;-c\leq
N_\nu\leq c,\e_\nu\leq c\right)\nonumber \\
&= \left[\inf_{\eta(X_\nu)} {\ex}\left(|\eta
-N_\nu|^p;-c\leq
N_\nu\leq c\Big|\e_\nu\leq c\right)\right]\pr(\e_\nu\leq c)\,.
\end{align}
Now,  $\ex \e_\nu$ can be upperbounded by  a constant $0\leq k<\infty$ that is
independent of the barrier level at time $\nu$, i.e., $\ell-sd(1-\varepsilon)$
(see \cite[Equation $(2)$]{Mog1}). Hence,  $$\pr(\e_\nu\leq c)\geq 1-k/c$$ by
Markov inequality. Therefore, for any fixed $0<\varepsilon<1$, $c$ large enough so
that 
\begin{align}\label{kc}
k/c\leq \varepsilon
\end{align} and
$d$ large enough so that \eqref{condc} holds, from \eqref{boulgo} we have
 \begin{align*}
&\frac{1}{(1-\varepsilon)}\frac{s^p}{({d(1-\varepsilon)})^{p/2}}\inf_{\eta}{\ex}|\eta
-\tl|^p\nonumber\\
&\geq  \inf_{ \eta(X_\nu)} {\ex}\left(|\eta
-N_\nu|^p;-c\leq
N_\nu\leq c\Big|\e_\nu\leq c\right)\,.
\end{align*}
For a fixed value of $\e_\nu$, 
$N_\nu\overset{\text{d}}{\longrightarrow}N$ by Claim iii. of Lemma \ref{lemma}
and by the strong Markov property of $X$ at time $\nu$.
Hence, $N_\nu\overset{\text{d}}{\longrightarrow}N$ uniformly over $\{\e_\nu\leq
c\}$.  Therefore, taking $\liminf_{d\to \infty}$ on both sides of the above
inequality we get
 \begin{align}\label{gou}
\liminf_{d\to \infty}\frac{1}{(1-\varepsilon)}\frac{s^p}{({d(1-\varepsilon)})^{p/2}}&\inf_{\eta}{\ex}|\eta
-\tl|^p\nonumber\\
&\geq  \inf_{ e} {\ex}\left(|e
-N|^p;-c\leq
N\leq c\right)\nonumber\\
&\geq  {\ex}\left(|N|^p;-c\leq
N\leq c\right)
\end{align}
where the infimum on the right-hand side of the second inequality is over constant
estimators, and where the last inequality follows from the symmetry and
monotonicity of the probability density function of $N$ around zero.

Since the above inequality holds for arbitrary $0<\varepsilon<1$ and
$c>0$ such that \eqref{kc} is satisfied, by letting $c=c(\varepsilon)=k/\varepsilon$ and
by taking $\varepsilon \to 0$ on both sides of \eqref{gou} yields
 \begin{align*}
\liminf_{d\to \infty}\frac{s^p}{{d}^{p/2}}&\inf_{\eta}{\ex}|\eta
-\tl|^p\geq  {\ex}|N|^p\,,
\end{align*}
implying that
 \begin{align*}
&\inf_{\eta}{\ex}|\eta
-\tl|^p\geq (1+o(1))\frac{{d}^{p/2}}{s^p} {\ex}|N|^p\,,
\end{align*}
as $d\to \infty$ while $\ell\geq s\cdot d$.

 \noindent{{{\textbf{Achievability:}}}} 
Let $\ell\geq s\cdot d$ and define $$\eta^{*}_d\defeq \inf\{t\geq 0: Y_t\geq \ell-s\cdot
d\}\,,$$ $$\xi\defeq \inf\{t\geq 0:
X_t\geq \ell-s\cdot d\}\,,$$  and $$\Delta\defeq \inf\{t\geq 0: X_{t+\eta}\geq
\ell\}\,.$$
These definitions imply that
$$\eta^{*}_d=\xi+d\,,$$ 
and
 $$\tau_\ell=\xi+\Delta\,.$$
Further, define $$\Delta_o\defeq \inf\{t\geq 0: X_{t+\xi}-X_\eta\geq sd\}\,.$$ Notice that if there were no
barrier overshoot at time $\xi$, then $X_\xi=\ell-s\cdot d$, and so $\Delta_o$ would be
equal to $\Delta$. 

It follows that
\begin{align}\label{p1}
\ex|\eta^{*}_d-\tau_\ell|^p&=\ex|\Delta-d|^p\nonumber \\
&\leq\left
[(\ex|\Delta_o-d|^p)^{1/p}+(\ex|\Delta_o-\Delta|^p)^{1/p}\right]^{p}\nonumber
\\
&=\left
[(\ex|\Delta_o-d|^p)^{1/p}+(\ex\tau_{\e_\xi}^p)^{1/p}\right]^{p}
\end{align}
where $$\e_\xi\defeq X_{\xi}-(\ell-s\cdot d)$$ denotes the excess at
time $\xi$. The first inequality in \eqref{p1} follows from the
triangle inequality and the second inequality
follows from the strong Markov property of $X$ at
time~$\xi$.

From Claim iii. of Lemma~\ref{lemma} and the strong Markov property of $X$ at
time~$\xi$,
\begin{align}\label{loubl}
\ex|\Delta_o-d|^p=(1+o(1))\frac{d^{p/2}}{s^p}\ex|N|^p
\end{align}
as $d\to \infty$.

Assume that $\ex {\tau_{\e_\xi}}^p$ can be upper bounded by a finite constant that does not depend
on $d$. Then, from
\eqref{p1} and \eqref{loubl} we get
$$\ex|\eta^{*}_d-\tau_\ell|^p\leq (1+o(1))\frac{d^{p/2}}{s^p}\ex|N|^p$$
as $d\to \infty$ while $\ell\geq s\cdot d$,
yielding the desired result.

As we now show, the fact that $\ex {\tau_{\e_\xi}}^p$ can be
upper bounded by a finite constant that does not
depend on $d$ essentially follows from
\cite[Equation (2)]{Mog1} which states that $\ex
{{\e_\xi}}^p$ can be upper bounded by a finite
constant that does not depend on the barrier level
at time $\eta$. For notational convenience, we
drop the subscript $\xi$ and write $\e$ in place of
$\e_\xi$.

If the
barrier level at time $\eta$, i.e., $(\ell-s\cdot
d)$, is bounded in the limit $d\to \infty$, i.e.,
if $\limsup_{d\to \infty}(\ell-s\cdot d)<\infty$,
then clearly $\ex {\tau_{\e}}^p$ can be
upper bounded by a finite constant that does not
depend on $d$.

Now, suppose that $\lim_{d\to
\infty}(\ell-s\cdot d)=\infty$, and suppose, by
contradiction, that $\ex {\tau_{\e}}^p\to
\infty$. We start with $p=1$.

By
Claim ii. of Lemma~\ref{lemma} we have 
\begin{align}\label{repi}
\tau_{\e}=\frac{\e}{s}+\left(\frac{\e}{s^3}\right)^{1/2}\tilde{N}_{\e}
\end{align}
where $\tilde{N}_{\e}\to N$ in distribution,
uniformly over $\{\e\geq k\}$, as $k\to \infty$.
Using this,
\begin{align}\label{repi2}
\ex {\tau_{\e}}&\leq \ex ({\tau_{k}}; \e\leq
k) +\ex ({\tau_{\e}}; \e\geq
k)\nonumber \\
&\leq \ex({\tau_k})+ \ex
({\e}/s+[{\e}/s^3]^{1/2}\tilde{N}_{\e};\e\geq k)\nonumber \\
&\leq \ex({\tau_k})+
(1/s)\ex{\e}+\ex(\tilde{N}_{\e}[{\e}/s^3]^{1/2})\nonumber \\
&\leq \ex({\tau_k})   + 
(1/s)\ex{\e}+s^{-3/2}[
(\ex{\e})\ex(\tilde{N}_{\e})^2]^{1/2}\nonumber\\
&\leq \ex({\tau_k})   + 
(1/s)\ex{\e}+s^{-3/2}[
(\ex{\e})(2\ex{N}^2)]^{1/2}\,.
\end{align}
The first inequality holds since $\tau_\ell\geq
\tau_{\ell'}$ for $\ell\geq \ell'$. The second
inequality follows from \eqref{repi}. The fourth inequality holds by
Cauchy-Schwartz inequality. The last inequality
holds by \eqref{repi} for $k$ large enough.

From \eqref{repi2}, if $\ex {\tau_{\e}}\to
\infty$ then $\ex \e\to \infty$, a contradiction
since \cite[Equation (2)]{Mog1} says that 
$\ex \e$ admits a finite upper bound that does
not depend on the barrier level. Hence,  $\ex {\tau_{\e}}\to
\infty$ can be upper bounded by a finite constant that does not depend on $d$.

For $p>2$, a similar argument as above shows that $\ex
\tau_\e^p<\infty$. In particular, a similar computation as in \eqref{repi} holds, with the addition of a
triangle inequality for the second inequality in \eqref{repi} to get
$$\ex(\tau_\e^p;\e\geq k)\leq \big(
(1/s)(\ex{\e^p})^{1/p}+(\ex(\tilde{N}_{\e}^p[{\e}/s^{3}]^{p/2}))^{1/p}\big)^{p}\,.$$
This shows for any $\ell=\ell(d)\geq s\cdot d$, $\limsup_{d\to \infty}\ex
\tau_\e^p<\infty$, yielding the desired result.
\subsection{Proof of Theorem~\ref{prop2}}
Fix $p\geq 1/2$. Suppose for the moment that a stopping time $\eta$ on $Y$ that satisfies $\pr
(\eta<\tau_\ell+d)>0$ also satisfies
\begin{align}\label{tof}
\ex (|\eta-\tau_\ell|^p|Y_\eta, \eta<\tau_\ell+d)=\infty\,.
\end{align}
Hence, if $\eta$ satisfies $\ex
|\eta-\tau_\ell|^p<\infty$, then necessarily $$\pr(\eta\geq
\tau_\ell+d)=1\,.$$
From this equality if follows that
\begin{align*}
\inf_\eta\ex |\eta-\tau_\ell|^p&=\inf_{\eta:\pr(\eta\geq \tau_\ell+d)=1}\ex
|\eta-\tau_\ell|^p\nonumber \\
&\geq d^p\nonumber \\
&=\ex|\eta^{*}_d-\tau_\ell|^p
\end{align*}
where $\eta^{*}_d=\inf\{t\geq 0: Y_t\geq \ell\}$. Therefore we have the desired
result
$$\inf_\eta\ex |\eta-\tau_\ell|^p=d^p=\ex|\eta^{*}_d-\tau_\ell|^p\,.$$

We prove \eqref{tof} assuming $\pr
(\eta<\tau_\ell+d)>0$. Equivalently, we show that for
any stopping rule $\eta$ over $X$ (instead of $Y$) such that
$\pr(\eta<\tau_\ell)>0$, necessarily we have
\begin{align}\label{tof1}
\ex (|\eta-\tau_\ell|^p|X_\eta,\eta<\tau_\ell)=\infty\,.
\end{align}

Given $X_\eta=\ell -h$, for some arbitrarily fixed $h>0$, let $\{B_{t}\}_{t \geq
0}$ be the continuous time version of $X$ starting at time $\eta$, i.e.,  $\{B_{t}\}_{t \geq 0}$ is a standard Wiener process starting at
time $\eta$ at level $B_{0}= \ell-h$ and such that $B_t=X_{\eta+t}$ for
$t=0,1,2,\ldots$.

Let
$$
\tilde{\tau_h} \defeq \inf \{t\geq 0: B_{t} =\ell\}.
$$
Suppose $\eta < \tau_\ell$. Since $\tilde{\tau_h} \leq \tau_\ell-\eta$, had we proved that
${\ex}\tilde{\tau_h}^p = \infty$, \eqref{tof1} would
hold.

From the reflection principle 
$$
{\pr}(\tilde{\tau}_h \leq t) = 2{\pr}(B_{t} \geq h) =
2Q\left(\frac{h}{\sqrt{t}}\right) \qquad h>0,\: t > 0\,,
$$
where
$Q(x)=(1/\sqrt{2\pi})\int_x^\infty\exp(-x^2/2)dx$.
Hence,
\begin{align*}
{\ex}\tilde{\tau}_h^p &= 2\int\limits_{0}^{\infty}
t^p dQ\left(\frac{h}{\sqrt{t}}\right) \\
&=
\frac{h}{\sqrt{2\pi}}\int\limits_{0}^{\infty}\frac{t^p}{t^{3/2}}
\,e^{-h^2/2t}dt \\
&> \frac{he^{-h/2}}{\sqrt{2\pi}}
\int\limits_{h}^{\infty}\frac{t^p}{t^{3/2}}\,dt.
\end{align*}
Therefore, if $p\geq 1/2$, then ${\ex}\tilde{\tau}_h^p =
\infty$, yielding the desired result.\hfill{$\blacksquare$}
\subsection{Proof of Lemma~\ref{lemma}}
 \noindent{{{\textbf{Claim i.}}}}  
For any real constant $q$, $S_t=\sum_{i=1}^t Z_i$ satisfies
$$
\ex\left[e^{q S_{t+1}}\big|S_{1},\ldots,S_{t}\right] =
e^{q S_{t} + q s + q^{2}\sigma^2/2}
$$
which can readily be checked by direct computation. 

Hence, letting 
$$M_{t} = e^{q S_{t} - r t} \qquad t\geq 1$$ where
$r$ is an arbitrary constant, we get
$$
\ex\left[M_{t+1}\big|M_{1},\ldots,M_{t}\right] = M_{t}
e^{q s + q^{2}\sigma^2/2 - r} \qquad t\geq 1.
$$
Let us set $r = q s + q^{2}\sigma^2/2$ so that 
$$ M_{t}=e^{q S_{t} - (q s +
q^{2}\sigma^2/2)t}\qquad t\geq 1$$ is a  martingale, and introduce the stopping time
$$
\underline{\tau_\ell} = \min\{\lceil k\rceil, \tau_\ell\}
$$
where $k >0$ is an arbitrary constant. It follows that
\begin{align*}
1 &= \ex M_{1} \nonumber 
\\&= \ex M_{\underline{\tau_\ell}} \nonumber\\
&\geq \ex [M_{\underline{\tau_\ell}}; \underline{\tau_\ell} < k]\nonumber\\
&\geq
e^{q \ell - (q s + q^{2}\sigma^2/2)k}\pr\left(\underline{\tau_\ell} <
k\right)  \quad  \quad q
\geq 0 \nonumber\\
&=
e^{q \ell - (q s + q^{2}\sigma^2/2)k}\pr\left( \tau_\ell < k\right)\,,
\end{align*}
where the second equality follows from Doob's
stopping theorem and where the second
inequality is valid for $q\geq 0$ since $S_{\tau_\ell}\geq \ell$
and $\underline{\tau_\ell}\leq n$.

It follows that
\begin{align}\label{eq:fina}
\pr\left(\tau_\ell < k\right)\leq e^{-q \ell + (q s +
q^{2}\sigma^2/2)k}\qquad q\geq 0\,.
\end{align}
 Minimizing the right-hand side
of \eqref{eq:fina} over $q\geq 0$ gives
\begin{align}\label{eq:epf}
\pr\left(\tau_\ell < k\right)\leq e^{-(\ell-sk)^2/2\sigma^2k}\,,
\end{align}
which is obtained for $q=q(k)=(\ell-sk)/\sigma^2 k$. Note that
this bound is valid for $k\leq \ell/s$ since $q$ should be nonnegative. By
assumption $k>0$, so inequality \eqref{stat2d}
follows from \eqref{eq:epf} by letting $k=\ell/s -z$, $0\leq z< \ell/s$.

Inequality \eqref{stat2a} follows from Chernoff bound. 

\noindent{{{\textbf{Claim ii.}}}}  
Using Claim i. and letting $u = \ell/s$, we have
\begin{align}\label{integr}
\ex\left|\tau_\ell - \ell/s\right|^{p}  
&= \int\limits_{0}^{\infty}\pr(|\tau_\ell - u| \geq z)
d(z^{p})\nonumber\\
 &\leq  2
 \int\limits_{0}^{\infty}e^{-s^{2}z^{2}/(2\sigma^2( u+z))}d(z^{p})\nonumber\\
  &\leq
2 \int\limits_{0}^{\infty}e^{-s^{2}z^{2}/(4\sigma^2\max\{u,z\})}d(z^{p})\nonumber\\
&= 2 (I_1+I_2)
\end{align}
where the first inequality follows from Claim i. and where
\begin{align*}
I_1&\defeq
\int\limits_{0}^{u}e^{-s^{3}z^{2}/(4\sigma^2\ell)}d(z^{p})\\
I_2&\defeq\int\limits_{u}^{\infty}e^{-s^{2}z/(4\sigma^2)}d(z^{p})\,.
\end{align*}

For $I_1$,  the change of variable 
$$
z = \left(\frac{4\sigma^2\ell}{s^{3}}\right)^{1/2}v^{1/p},
$$
yields
\begin{align}\label{i1}
I_{1} &= \left(\frac{4\sigma^2\ell}{s^{3}}\right)^{p/2}
\int\limits_{0}^{[s^{3}u^{2}/(4\sigma^2\ell)]^{p/2}}e^{-v^{2/p}}dv \nonumber\\
&\leq
\left(\frac{4\sigma^2\ell}{s^{3}}\right)^{p/2}
\int\limits_{0}^{\infty}e^{-v^{2/p}}dv \nonumber \\
&=k_1\ell^{p/2} \qquad p > 0.
\end{align}
where $0<k_1<\infty$ is a constant that depends on $s$, $p$, and $\sigma^2$.
 
For $I_{2}$, the change of variables $z = v^{1/p}$ and
$v=ts^{-2p}$ yield
\begin{align}\label{i2}
I_{2} &=
\int\limits_{u^{p}}^{\infty}e^{-s^{2}v^{1/p}/(4\sigma^2)}dv
\nonumber\\
&\leq
e^{-s^{2}u/(8\sigma^2)}\int\limits_{u^{p}}^{\infty}e^{-s^{2}v^{1/p}/(8\sigma^2)}dv\nonumber\\
& =
e^{-s^{2}u/(8\sigma^2)}s^{-2p}\int\limits_{(s^{2}u)^{p}}^{\infty}
e^{-t^{1/p}/(8\sigma^2)}dt \nonumber\\
& \leq
e^{-s^{2}u/(8\sigma^2)}s^{-2p}\int\limits_{0}^{\infty}e^{-t^{1/p}/(8\sigma^2)}dt
\nonumber\\
& \leq
s^{-2p}\int\limits_{0}^{\infty}e^{-t^{1/p}/(8\sigma^2)}dt
\nonumber\\
&=k_2
\end{align}
where $0<k_2<\infty$ is a constant that depends on $s$, $p$, and $\sigma^2$.
From \eqref{integr}, \eqref{i1}, and \eqref{i2}
\begin{equation*}
\ex |\tau_\ell-\ell/s|^p \leq k_3(k_4+\ell^{p/2})
\end{equation*}
for some constants $k_3$ and $k_4$ that depend on $s$, $p$, and $\sigma^2$. This
yields the desired result.

\noindent{{{\textbf{Claim iii:}}}}  See \cite[Theorem 2.5]{gut}. \hfill{$\blacksquare$}

\section*{Acknowledgments} The authors are grateful to the reviewers and to the
Associate Editor for their insightful and detailed comments on the manuscript, and for questionning
the non-asymptotic behavior of $\etans$ which prompted the investigation
of the complementary stopping rule $\etas$. 

The authors are also indebted to Milad
Sefidgaran for many helpful discussions.

\section*{Appendix}
\subsubsection*{Simulation - noisy observations}
To numerically evaluate \eqref{ler} for $\eta=\{\etas,\etans,\ell/s\}$, for each
given value of $\ell$
we generated $n$ samples of $(X,Y)$, and
computed the corresponding empirical sums $$s_n=\frac{1}{n}\sum_{i=1}^n |\eta(i)-\tl(i)|\qquad
\eta\in\{\etas,\etans,\ell/s\}$$ where
$(\eta(i),\tl(i))$ is the value of $(\eta,\tl)$ for the $i\/$-th sample of
$(X,Y)$.\footnote{To be precise, we sequentially generated 
$(X_1,Y_1),(X_2,Y_2),\ldots$, until
both $\tl$ and $\eta$ had stopped. So the generated samples $(X,Y)$'s are of
variable length.}


Letting $$C_1(\ell, p)=C_1(\ell,\varepsilon, s, p)$$ be the constant defined in
Theorem~\ref{th:main} with $\varepsilon=.5$ and $s=10$,  Chebyshev's inequality
gives the sufficient condition
on the number of samples $n$
\begin{align}\label{cheby}
n\geq  \frac{\text{Var}(\eta-\tl)}{ \delta^3\cdot  C_1^2(\ell,1) }
\end{align}
in order to have
\begin{align}\label{ch2}
\pr\left(\frac{1}{C_1(\ell,1)}\big|s_n-\ex|\eta-\tl|\big|\leq \delta \right)\geq 1-\delta \,.
\end{align}
To use \eqref{cheby}, we need to evaluate $\text{Var}(\eta-\tl)$. To do this,
observe that $\ex \eta\approx \ex \tau_\ell\approx \ell/s$ for $\eta\in
\{\etans,\etas,\ell/s\}$ (these approximations become
equalities if we ignore overshoot). So we have 
\begin{align}
\text{Var}(\eta-\tl)&\approx \ex|\eta-\tl|^2\nonumber \\
&=\left\{ 
\begin{array}{ll}
(1+o(1))C_1(\ell,2) & \eta=\etas\quad \text{or}\quad \eta=\etans\\
(1+o(1))(\ell/s^{3})&  \eta =\ell/s
\end{array}
\right.\label{lbb}
\end{align}
where the equality follows from Theorem~\ref{th:main} and \eqref{etals}. Combining \eqref{cheby}
together with \eqref{lbb} gives
\begin{align}\label{wap}
n&\gtrsim \frac{\pi}{2\cdot\delta^3} \qquad\text{for}\quad \eta=\etas\quad
\text{or}\quad \eta=\etans\nonumber \\
n&\gtrsim \frac{5\cdot\pi }{2\cdot \delta^3} \qquad\text{for}\quad \eta=\ell/s
\end{align}
as a reasonable condition on $n$ for \eqref{ch2} to hold. In Fig.~\ref{fig},
$n=10,000$ which guarantees roughly
$\delta=.05$ for $\eta=\etas$ or $\eta=\etans$ and $\delta=.1$ for $\eta=\ell/s$.

 Finally note that, for small values of $\ell$, the contribution due to overshoot
cannot be neglected and Theorem~\ref{th:main} is loose. So in this regime
the bounds \eqref{wap} must be taken with a grain of
salt.

\subsubsection*{Simulation - delayed observations}
We proceeded similarly as in the previous section. We generated $n$ samples
$X$, computed the corresponding empirical sums $s_n$ with
$\eta=\eta^{*}$, and finally used Chebyshev's related inequality \eqref{cheby}
with $\text{Var}(\eta-\tau_\ell)=C_2(d,s,2)$ and $C_1(\ell,1)$ replaced by
$C_2(d,s,1)$ to obtain
\begin{align}\label{cheby2}
n\geq  \frac{C_2(d,s,2)}{ \delta^3\cdot  C_2^2(\ell,s,1) }=\frac{\pi}{2\delta^3}
\end{align}
as a reasonable condition on $n$ to achieve $\delta$ precision. 
In Fig.~\ref{fig2}, $n=100,000$ which guarantees
a precision of $\delta=.03$.

\section*{Biographies}

{\bf Marat V. Burnashev} was born in Tashkent, USSR, on January 6,
1947. He graduated and received Ph.D. degree in mathematics from the
Moscow Institute of Physics and Technology in 1971 and 1974, 
respectively. He received the Full Doctor degree in mathematics from the
Mathematical (Steklov) Institute of the USSR Academy of Sciences
in 1985.

Since 1974, he has been with the Institute for Information Transmission
Problems of the Russian Academy of Sciences. He has also held
visiting appointments at several universities in the USA, Canada,
Japan, Germany, France, etc.

His research interests include information theory, probability
theory, and mathematical statistics.

{\bf Aslan Tchamkerten} received the Engineer Physicist Diploma in 2000 and
the Ph.D. degree in Communications in 2005, both from the Ecole
Polytechnique F\'ed\'erale de Lausanne (EPFL), Switzerland. Between 2005 and
2008, he was a Postdoctoral Associate in the Department of Electrical
Engineering and Computer Science, Massachusetts Institute of Technology
(MIT), Cambridge. In 2008 he joined Telecom ParisTech (ex. Ecole
Nationale Sup\'erieure des T\'l\'ecommunications,ENST), Paris, France, where he
is currently Associate Professor. In 2009, he won
a junior excellence chair grant from the French National Research Agency
(ANR). His research interests are in Information Theory, Applied Statistics,
and Algorithms.



\bibliography{../../../common_files/bibiog}

\begin{thebibliography}{10}

\bibitem{BN2}
M.~Basseville and I.~Nikiforov.
\newblock {\em Detection of abrupt changes: theory and application}.
\newblock Prentice-Hall, 1993.

\bibitem{BT}
M.~V. Burnashev and A.~Tchamkerten.
\newblock Tracking a {G}aussian random walk first-passage time through noisy
  observations.
\newblock {\em accepted in Ann. App. Prob.}, 2010.

\bibitem{CLRS}
T.~H. Cormen, C.~E. Leiserson, R.~L. Rivest, and C.~Stein.
\newblock {\em Introduction to Algorithms, 2nd edition}.
\newblock {MIT} Press, McGraw-Hill Book Company, 2000.

\bibitem{DeG}
M.~.H. DeGroot.
\newblock {\em Optimal Statistical Decisions}.
\newblock Wiley, Hoboken (NJ), 2004.

\bibitem{gut}
A.~Gut.
\newblock On the moments and limit distributions of some first passage times.
\newblock {\em Ann. Prob.}, 2(2):277--308, 1974.

\bibitem{Lai1}
T.Z. Lai.
\newblock Information bounds and quick detection of parameter changes in
  stochastic systems.
\newblock {\em IEEE Trans.~Inform.~Th.}, 44:2917--2929, November 1998.

\bibitem{Mog1}
A.A. Mogulskii.
\newblock Absolute estimates for moments of certain boundary functionals.
\newblock {\em Th. Prob. Appl.}, 18(2):350--357, 1973.

\bibitem{Mo2}
G.V. Moustakides.
\newblock Sequential change detection revisited.
\newblock {\em Ann. Statist.}, 36(2):787--807, 1988.

\bibitem{NT}
U.~Niesen and A.~Tchamkerten.
\newblock Tracking stopping times through noisy observations.
\newblock {\em IEEE Trans.~Inform.~Th.}, 55(1):422--432, January 2009.

\bibitem{PH}
H.V. Poor and O.~Hadjiliadis.
\newblock {\em Quickest detection}.
\newblock Cambridge, New York, 2009.

\bibitem{Shi3}
A.~N. Shiryaev.
\newblock On optimum methods in quickest detection problems.
\newblock {\em Th. Prob. and its App.}, 8(1):22--46, 1963.

\bibitem{Shi}
A.~N. Shiryayev.
\newblock {\em Optimal Stopping rules}.
\newblock Springer-Verlag, 1978.

\bibitem{TM}
A.~G. Tartakovsky and G.~Moustakides.
\newblock State-of-the-art in bayesian changepoint detection.
\newblock {\em Seq. Analysis}, 29(2):125--145, 2010.

\bibitem{Y}
B.~Yakir.
\newblock Optimal detection of a change in distribution when the observations
  form a {M}arkov chain with a finite state space.
\newblock In {\em Change-point problems}, volume~23, pages 346--358. Institute
  of Mathematical Statistics, Lecture Notes, Monograph Series, 1994.

\end{thebibliography}


\end{document}